\documentclass[a4paper,UKenglish,cleveref, autoref, thm-restate]{lipics-v2021}
\usepackage{graphicx}
\usepackage{hyperref}
\usepackage{subfiles}
\usepackage{amsmath}
\usepackage{cleveref}
\usepackage{dsfont}
\usepackage{ifsym}
\usepackage{comment}
\usepackage[dvipsnames]{xcolor}
\usepackage{boxedminipage}
\usepackage[ruled,vlined]{algorithm2e}

\RestyleAlgo{ruled}

\SetKwComment{Comment}{/* }{ */}

\title{Exact Matching and the Top-k Perfect Matching Problem} 

\author{Nicolas {El Maalouly}}{Department of Computer Science, ETH Z\"{u}rich, Switzerland }{nicolas.elmaalouly@inf.ethz.ch}{0000-0002-1037-0203}{}
\author{Lasse {Wulf}}{Institute of Discrete Mathematics, TU Graz, Austria }{wulf@math.tugraz.at}{0000-0001-7139-4092}{Supported by the Austrian Science Fund (FWF): W1230.}

\authorrunning{N. El Maalouly, L. Wulf} 

\Copyright{Nicolas El Maalouly, Lasse Wulf} 



\EventEditors{John Q. Open and Joan R. Access}
\EventNoEds{2}
\EventLongTitle{42nd Conference on Very Important Topics (CVIT 2016)}
\EventShortTitle{CVIT 2016}
\EventAcronym{CVIT}
\EventYear{2016}
\EventDate{December 24--27, 2016}
\EventLocation{Little Whinging, United Kingdom}
\EventLogo{}
\SeriesVolume{42}
\ArticleNo{23}

\begin{document}
\nolinenumbers
\maketitle

\begin{abstract}

The aim of this note is to provide a reduction of the Exact Matching problem to the Top-$k$ Perfect Matching Problem. Together with earlier work by El Maalouly, this shows that the two problems are polynomial-time equivalent. 

The Exact Matching Problem is a well-known 40 years old problem for which a randomized, but no deterministic poly-time algorithm has been discovered. The Top-$k$ Perfect Matching Problem is the problem of finding a perfect matching which maximizes the total weight of the $k$ heaviest edges contained in it. 
\end{abstract}

\section{Reduction}

Exact Matching (EM), defined in 1982 by Papadimitriou and Yannakakis~\cite{papadimitriou1982complexity}, is one of only few natural problems which is known to be solvable in randomized polynomial time, but for which no deterministic poly-time algorithm is known so far.
\vspace{5pt}

\noindent\vspace{5pt}\begin{boxedminipage}{\textwidth}
\textsc{Exact Matching (EM)}

\textbf{Input:} A graph $G$, where every edge is colored blue or red, and an integer $k$.

\textbf{Task:} Decide whether there exists a perfect matching $M$ in $G$ with exactly $k$ red edges.

\end{boxedminipage}

Given a weight function $w : E \rightarrow \mathbb{R}$ which assigns a weight $w(e)$ to every edge of a graph, and given a subset $F \subseteq E$ of the edges, we order the elements of $F$ by their weight. For a given integer $k$ we let $w^k(F)$ denote the sum of the weight of the $k$ heaviest elements in $F$. The function $w^k(\cdot)$ is called the \emph{top-$k$} weight function.

We show that EM can be reduced (in deterministic polynomial time) to the following optimization problem defined and studied in \cite{elmaalouly2022exacttopk}.

\vspace{5pt}
\noindent\vspace{5pt}\begin{boxedminipage}{\textwidth}

\textsc{Top-$k$ Perfect Matching (TkPM)}

\textbf{Input:} A weighted graph $G$ and integer $k$.

\textbf{Task:} Find a perfect matching $M$ in $G$ maximizing the top-$k$ weight function $w^k(M)$.

\end{boxedminipage}

\begin{lemma}
$EM \leq_p TkPM$, even if the edge weights in the TkPM instance are bounded by a constant.
\end{lemma}

\begin{proof}
    Consider an instance of EM, which is given by a graph $G := (V,E)$ and a red-blue coloring of its edges, and an integer $k$. We describe how to obtain in polynomial time an instance of TkPM given by a graph $G' = (V',E')$ and a weight function $w : E' \rightarrow \mathbb{R}$ and an integer $k'$. We start with $G$ and subdivide all edges four times, i.e. every edge is replaced by a path of length $5$. For every edge $e \in E$, let $P_e \subseteq G'$ be the path replacing it. In addition to this, we add $2k$ new vertices and $k$ independent edges forming a perfect matching of these $2k$ vertices to the graph $G'$. Let $E_k$ be the set of these $k$ edges. This completes the description of $G'$. Observe that any perfect matching in $G'$ must contain the set $E_k$.
    
    For the weight function $w : E' \rightarrow \mathbb{R}$, we let $w(e) = 2$ for all $e \in E_k$. For all the edges on some path $P_e$, we distinguish the case whether $e$ is colored red or blue in $G$. If $e$ is blue, all edges of $P_e$ get weight $0$. If $e$ is red, the middle edge of $P_e$ gets weight $3$, the two edges adjacent to it get weight $2$ and the two remaining outer edges get weight 0.
    
    Finally, let $k' := 2|R(G)|$ where $|R(G)|$ is the number of red edges in $G$. This completes our description of the TkPM instance.
    Let $\alpha := 4|R(G)| + k$. We now claim that there is a PM in $G$ with exactly $k$ red edges if and only if there is a PM $M'$ in $G'$ with $w^{k'}(M') \geq \alpha$.
    

    To show this, observe that there is a one-to-one correspondence between perfect matchings in $G$ and perfect matchings in $G'$ where an edge is in the perfect matching of $G$ if and only if the middle edge of $P_e$ is in the perfect matching of $G'$. Let $M \subseteq G$ and $M' \subseteq G'$ be two such perfect matchings which correspond to each other. 
    Let $r=|R(M)|$ be the number of red edges in the matching $M$. 
    Note that 
    \[ w^{k'}(M') \leq 3 r + 2 (k'-r) =  4|R(G)| + r. \] 
    This inequality is due to the fact that the maximum weight of an edge is 3, but every edge of weight 3 is a middle edge of some path $P_e$ where $e \in E(G)$ is colored red. Therefore the $k'$ heaviest edges in $M'$ can contain at most $r$ edges of weight 3 and at most $k' - r$ edges of weight 2.
    
    This shows that for the matching $M'$ to achieve $w^{k'}(M') \geq \alpha$, we need at least $k$ red edges in $M$.  Finally suppose that $M$ has at least $k$ red edges, that is $r \geq k$. Consider all the edges of non-zero weight in $M'$. These are exactly the edges of weight 3 corresponding to the middle of a path $P_e$ where $e$ is red and $e \in M$, and all pairs of edges of weight 2 corresponding to a path $P_e$ where $e$ is red and $e \not\in M$, and all the edges in $E_k$. (Observe that paths $P_e$ where $e$ is blue have weight 0).
    We count the number of non-zero weight edges and observe that this number is $r + 2(|R(G)| - r) + k = 2|R(G)| + k - r$. Using the assumption $r \geq k$ we have that the number of non-zero edges is smaller or equal to $k'$, so every non-zero edge is included in $w^{k'}(M')$. In total, we have
    $w^{k'}(M') = 3 \cdot r + 2k + 4 \cdot (|R(G)|-r) = 4|R(G)| -r + 2k.$ This number is equal to $\alpha$ in the case $r=k$ and smaller than $\alpha$ in the case $r > k$.
    
    We conclude that for all pairs $(M, M')$ of corresponding perfect matchings we have $w^{k'}(M') \geq \alpha$ if and only if $M$ has exactly $k$ red edges.
\end{proof}

In combination with the results of \cite{elmaalouly2022exacttopk}, we get that EM and TkPM are polynomially equivalent. Note that the reduction described there is not a Karp-reduction. 

\bibliography{references}


\end{document}